\documentclass[fleqn,usenatbib]{mnras}

\usepackage{newtxtext,newtxmath}
\usepackage[T1]{fontenc}

\DeclareRobustCommand{\VAN}[3]{#2}
\let\VANthebibliography\thebibliography
\def\thebibliography{\DeclareRobustCommand{\VAN}[3]{##3}\VANthebibliography}

\usepackage{graphicx}	% Including figure files
\usepackage{amsmath}	% Advanced maths commands
\usepackage{bm}
\usepackage{float}
\title[Low eccentricity exterior resonances]{Non-perturbative investigation of low eccentricity exterior mean motion resonances}

\author[Malhotra and Chen]{
Renu Malhotra,$^{1}$\thanks{Email: malhotra@arizona.edu}
Zherui Chen$^{2}$\thanks{E-mail: zr-chen20@mails.tsinghua.edu.cn}
\\
$^{1}$Lunar and Planetary Laboratory, The University of Arizona, Tucson, AZ 85721, USA\\
$^{2}$Mechanics, Xingjian College, Tsinghua University, Beijing 100084, China\\
}

\date{Accepted XXX. Received YYY; in original form ZZZ}

\pubyear{2022}

\begin{document}
\label{firstpage}
\pagerange{\pageref{firstpage}--\pageref{lastpage}}
\maketitle

\begin{abstract}
%This is a simple template for authors to write new MNRAS papers. The abstract should briefly describe the aims, methods, and main results of the paper. It should be a single paragraph not more than 250 words (200 words for Letters). No references should appear in the abstract.
%{\color{blue}[1-2 sentences on background and motivation, 1-2 sentences on methodology, 1-2 sentences on results, 1-2 sentences on implications]}
Mean motion resonances are important in the analysis and understanding of the dynamics of planetary systems. While perturbative approaches have been dominant in many previous studies, recent non-perturbative approaches have revealed novel properties in the low eccentricity regime for interior mean motion resonances of Jupiter in the fundamental model of the circular planar restricted three body model. Here we extend the non-perturbative investigation to exterior mean motion resonances in the low eccentricity regime (up to about 0.1) and for perturber mass in the range $\sim5\times10^{-5}$ to $1\times10^{-3}$ {(in units of the central mass)}. Our results demonstrate that first order exterior resonances have two branches at low eccentricity as well as low-eccentricity bridges connecting neighboring first order resonances. With increasing perturber mass, higher order resonances dissolve into chaos whereas low order resonances persist with larger widths in their radial extent but smaller azimuthal widths. For low order resonances, we also detect secondary resonances arising from small integer commensurabilities between resonant librations and the synodic frequency. These secondary resonances contribute significantly to generating the chaotic sea that typically occurs near mean motion resonances of higher-mass perturbers.   
\end{abstract}

\begin{keywords}
celestial mechanics -- three body problem -- orbital resonances -- chaos 
\end{keywords}

%%%%%%%%%%%%%%%%%%%%%%%%%%%%%%%%%%%%%%%%%%%%%%%%%%

%%%%%%%%%%%%%%%%% BODY OF PAPER %%%%%%%%%%%%%%%%%%

\section{Introduction}\label{s:intro}

%This is a simple template for authors to write new MNRAS papers. See \texttt{mnras\_sample.tex} for a more complex example, and \texttt{mnras\_guide.tex} for a full user guide.
%
%All papers should start with an Introduction section, which sets the work in context, cites relevant earlier studies in the field by \citet{Fournier1901}, and describes the problem the authors aim to solve \citep[e.g.][]{vanDijk1902}. Multiple citations can be joined in a simple way like \citet{deLaguarde1903, delaGuarde1904}.

Mean motion resonances are important for many aspects of the dynamics of planetary systems. A basic quantity about a mean motion resonance (henceforth abbreviated as MMR) is its width in terms of the range of the ratio, $n'/n$, of the mean motions of two objects {whose orbital periods are} close to a resonant ratio of two mutually prime integers, $j/(j+k)$. This width, which depends on the masses of the objects as well as the orbital eccentricities, is often reported as a range of the semimajor axis ratio, $a/a' = (n'/n)^\frac{2}{3}$. Within this resonance width the conjunctions of the two objects librate rather than circulate relative to the apsidal line of one of the bodies (or, sometimes, relative to their mutual nodal line); this libration is a consequence of the mutual gravity of the two objects. Many previous theoretical investigations have taken a perturbative approach; the reader is referred to the textbooks by \cite{Murray:1999SSD} and \cite{Morbidelli:2002Book}, as well as other recent literature \citep[e.g.][]{Deck:2013,Hadden:2019,Gallardo:2019,Lei:2020}. The perturbative approaches, especially analytical treatments, have been very useful in estimating the sizes of resonance zones, their scalings with parameters, as well as revealing many qualitative and quantitative properties of the dynamics near resonances  \citep[e.g.][]{Henrard:1983,Wisdom:1980}. Perturbative approaches have also been employed in modeling and analysis of observations of exo-planets whose low-eccentricity orbits are near first order MMRs \citep[e.g.,][]{Holman:2005,Lithwick:2012,Wu:2013}.

Unsurprisingly, perturbative approaches have limitations.
A recent investigation of the resonance widths in the planar circular restricted three body problem reported novel details of first order interior MMRs in the low eccentricity regime \citep{Malhotra:2020}; the novel details include the existence of a separatrix at zero eccentricity and ``low eccentricity resonant bridges" between neighboring interior first order resonances. These hitherto unknown features were found by making use of a non-perturbative approach that was developed for the investigation of interior resonance widths in the high eccentricity regime \citep{Wang:2017}. These features have subsequently been confirmed with higher order perturbative methods \citep{Lei:2020,Antoniadou:2021}; a brief review is given in \cite{Malhotra:2022a}. The same approach was used by \cite{Malhotra:2018a} and \cite{Lan:2019} to measure the widths of Neptune's exterior resonances in the moderate to high eccentricity regime; however, these authors did not investigate the low eccentricity regime of those exterior MMRs, below $e\approx 0.05$, because it presents certain numerical challenges.

Here we apply the non-perturbative approach to investigate the widths of low order exterior mean motion resonances, in the low eccentricity regime.  (The order of a $j:(j+k)$ MMR is the absolute value of the integer $k$, as the perturbation strength is of order $e^{|k|}$, where $e$ refers to the orbital eccentricity of the perturbed particle.) In addition to obtaining non-perturbative measurements of resonance widths, we are also interested to know if the zero eccentricity separatrix and the ``low eccentricity resonant bridges" can also be found near first order exterior MMRs. While we expect the answer is affirmative, we expect differences in detail because interior and exterior resonances have quantitative and qualitative asymmetries. For example, it is known that, unlike the interior 2:1 resonance, the exterior 1:2 resonance has a bifurcated libration center that leads to three libration zones --- a large-amplitude zone of librations surrounding a pair of so-called ``asymmetric" libration zones. This phenomenon is peculiar to exterior $1:j$ MMRs \citep[e.g.][]{Beauge:1994,Malhotra:1996,Murray-Clay:2005,Lan:2019}. 

The paper is organized as follows. The methodology is described briefly in Section~\ref{s:methodology}. The results are given in Section~\ref{s:results}. In Section~\ref{s:summary}, we summarize our findings and discuss broader implications.

\section{Methodology}\label{s:methodology}

The non-perturbative method makes use of Poincar\'e surfaces of section of the planar circular restricted three body problem.  The procedure is essentially the same as detailed in a recent sequence of papers \citep{Wang:2017,Malhotra:2018a,Lan:2019,Malhotra:2020}.  Here we provide a brief description of our implementation for exterior resonances in the regime of low eccentricity of the massless body. The equations of motion for the massless body are solved numerically with the adaptive step size seventh order Runge-Kutta method \citep{Fehlberg:1968}, with relative and absolute error tolerance of $10^{-12}$. We use natural units for the restricted three body problem, namely, the unit of mass is the total mass of the two primaries, $m_1+m_2$, the unit of length is the constant distance between them (the radius of their relative motion in a circular orbit), and the unit of time is their orbital period divided by $2\pi$. The equations of motion of the massless third body are expressed in the rotating reference frame of unit angular velocity whose origin is at the barycenter of the two primaries.  In this reference frame, the primaries are at fixed locations on the $x$-axis, at $(-\mu,0)$ and $(1-\mu,0)$, where $\mu=m_1/(m_1+m_2)$. The position of the massless body is recorded at every pericenter passage. The pericenter passage is identified by monitoring the length of the position vector of the massless body, and using interpolation when a pericenter passage is detected in two consecutive integration steps. The initial conditions are given in osculating orbital elements, which are then converted to the initial position and velocity in the rotating reference frame. Initial conditions are chosen as follows. For each surface-of-section, we initialize all trajectories such that they share a common value of the Jacobi constant, $C_J$. Initial semi-major axis $a$ and eccentricity $e$ are chosen in the neighborhood of the resonance of interest; the combination of $a$ and $e$ is constrained by the Jacobi constant. All trajectories also have zero initial mean anomaly (so they start at pericenter). The initial longitude of pericenter is different for different trajectories in the same surface-of-section so as to cover the full range of the angular elements in the resonance neighborhood.   In the work reported here, we investigated several different values of the mass ratio, $\mu$, in the planetary mass range $5.146\times10^{-5}$ to $1\times10^{-3}$. 

From the record of consecutive pericenter passages, we calculate the barycentric osculating orbital elements of the test particle and make plots of the Poincar\'e surfaces-of-section in $(\psi,a)$ and in $(e\cos\psi,e\sin\psi)$, where $a$ is the semimajor axis, $e$ is the eccentricity, and $\psi$ is the angular separation of the planet from the test particle at the latter's pericenter passage. In these surfaces-of-section, we first identify the resonant islands. For each resonant island, we measure the values $\psi_\mathrm{res}$, $a_\mathrm{res}$ and $e_\mathrm{res}$ at the centers of the resonant libration islands, and also measure the minimum and maximum values of $a$ in each libration island. The latter defines the width of the resonance with center at $(a_\mathrm{res},e_\mathrm{res})$.

For later reference, we note that the usual critical resonant angle for an exterior $j:(j+k)$ MMR is defined by
\begin{equation}
    \phi=(j+k)\lambda- j \lambda' - k\varpi ,
\label{e:phi}\end{equation} 
where $\lambda$ and $\varpi$ are the mean longitude and the longitude of pericenter of the particle, and $\lambda'$ is the mean longitude of the perturbing planet.
While $\phi$ is a continuous variable of time, when we compute its value at the time of pericenter passage of the  particle, it is an integer multiple of our stroboscopic variable $\psi$, i.e., $\phi = j\psi$. 

It is also useful to note that the extent of the stable resonant islands in terms of $a$ and $e$ relates to the radial extent of stable resonance libration, and the extent of the resonant islands in terms of $\psi$ measures the azimuthal extent of the stable librations. {The radial extent of the resonance libration refers to the libration of the perihelion distance in the range $a_\mathrm{min}(1-e_\mathrm{min})$ to $a_\mathrm{max}(1-e_\mathrm{max})$. Here the subscripts $\mathrm{min}$ and $\mathrm{max}$ refer to the minimum and maximum values of the parameters measured in a resonant island in the Poincar\'e section. [Note that the Jacobi constant enforces a corelation in the time variations of $a$ and $e$, so minimum and maximum perhelion distance is given as stated instead of the naive expectation that it would be $a_\mathrm{min}(1-e_\mathrm{max})$ and $a_\mathrm{max}(1-e_\mathrm{min})$, respectively.] The azimuthal extent of the libration is the range, $\psi_\mathrm{min}$ to $\psi_\mathrm{max}$, of the librations of the perihelion longitude in the rotating frame.}

The low eccentricity regime presents greater numerical challenges because the resonance libration zones in the surfaces of section are small and difficult to resolve for mass ratios $\mu$ in the planetary mass regime. Secondary resonances and chaotic layers crowd the surfaces of section, even for low order resonances, requiring greater effort and care in identifying the boundaries of stable libration zones. For exterior $1:j$ resonances, there is the additional complexity arising from bifurcated libration centers. This requires numerical experimentation with a much larger number of initial conditions and of finer numerical resolution. The investigation of the low eccentricity regime of exterior resonances is more time consuming in both human time and in computing time.

\section{Results}\label{s:results}

\begin{table}
\caption{
Resonance centers and boundaries of exterior MMRs. \\ This is a small sample of the tabulated measurements. The complete table is available in machine readable form in the online supplement.
}
\begin{center}
\begin{tabular}{cccccc}
\hline
$e_\mathrm{res}$ & $a_\mathrm{res}$ & $a_\mathrm{min}$ & $a_\mathrm{max}$ & $\psi_\mathrm{res}$ & MMR \\
\hline
 $\bm{\mu=9.53\times10^{-4}}$ &  &   &   &   &   \\
$0.0982$ &$2.0730$& $2.0622$& $2.0885$& $180^\circ$& 1:3\\
$0.0912$ &$2.0733$& $2.0639$& $2.0872$& $180^\circ$& 1:3\\
... & ... & ... & ... & ... & ...  \\
... & ... & ... & ... & ... & ...  \\
$ 0.0062$& $ 2.0725$& $2.0722 $& $2.0744 $& $180^\circ$& 1:3  \\
$0.0033$& $ 2.0726 $& $ 2.0724 $& $2.0731  $& $180^\circ$& 1:3  \\

$0.0964  $& $1.8393  $& $1.8305  $& $1.8539  $& $\pm 89.95^ \circ $& 2:5 \\
$0.0756  $& $ 1.8404 $& $1.8327 $& $ 1.8504 $& $\pm91.80^\circ $& 2:5 \\
... & ... & ... & ... & ... & ...  \\
... & ... & ... & ... & ... & ...  \\
$ 0.0114 $& $ 1.8411 $& $ 1.8373 $& $ 1.8442 $ &$\pm95.56^\circ $ & 2:5 \\
$ 0.0079 $& $1.8412  $& $ 1.8385 $& $ 1.8425$& $\pm98.65^ \circ $&2:5  \\

... & ... & ... & ... & ... & ...  \\
... & ... & ... & ... & ... & ...  \\

\hline

$\bm{\mu=5.146\times10^{-5}}$ &  &   &   &   &   \\
$0.0918 $& $2.0797 $& $2.0779 $& $2.0816 $& $180^\circ$&  1:3\\
 
... & ... & ... & ... & ... & ...  \\
... & ... & ... & ... & ... & ...  \\
$0.0076 $& $2.0798 $& $2.0795 $& $2.0802 $& $180 ^\circ$&  1:3\\
$0.0025$& $2.0798 $& $2.0798$& $2.0801$& $180 ^\circ$&  1:3\\
... & ... & ... & ... & ... & ...  \\
... & ... & ... & ... & ... & ...  \\
\noalign{(and so on)}\\
\hline
\end{tabular}
\end{center}
\label{t:table1}
\end{table}

Our main results consist of a table of measured values of $\psi_\mathrm{res}$, $e_\mathrm{res}$, $a_\mathrm{res}$, $a_\mathrm{min}$, $a_\mathrm{max}$. This tabulated data is available in the online article as an electronically readable file. A sample of this table is presented in Table~\ref{t:table1}. These results are based on measurements from many Poincar\'e surfaces-of-section. Below we  provide a selection of plots of the Poincar\'e surfaces-of-section and associated summary plots of the data, in order to illustrate and discuss the results. We have organized these plots as follows. 

\begin{figure*}
    \vskip-0.5truein
    \includegraphics[width=0.33\textwidth]{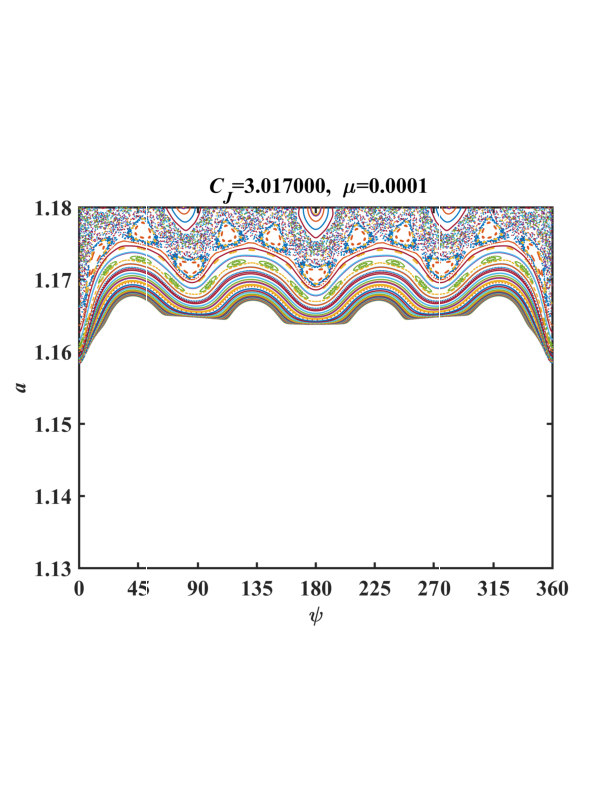}
 	\includegraphics[width=0.33\textwidth]{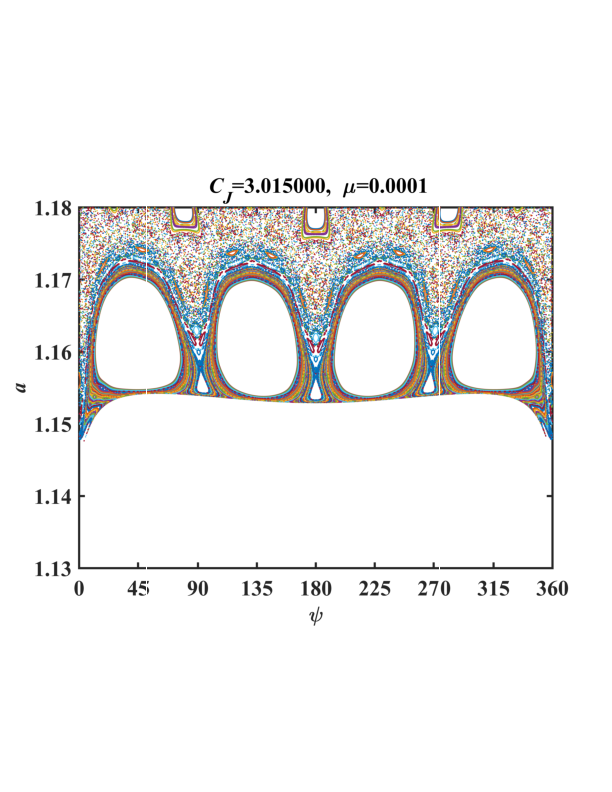}
  	\includegraphics[width=0.33\textwidth]{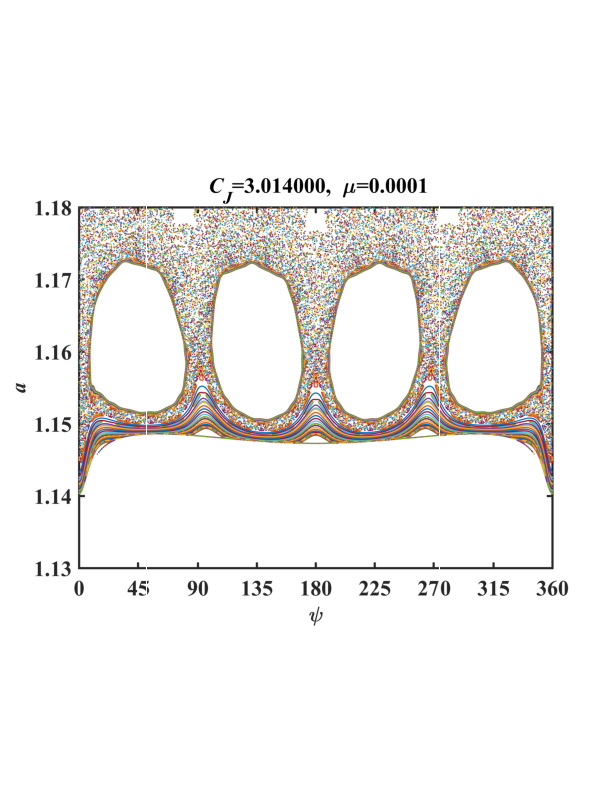}\\
      \vskip-1.1truein
    \includegraphics[width=0.33\textwidth]{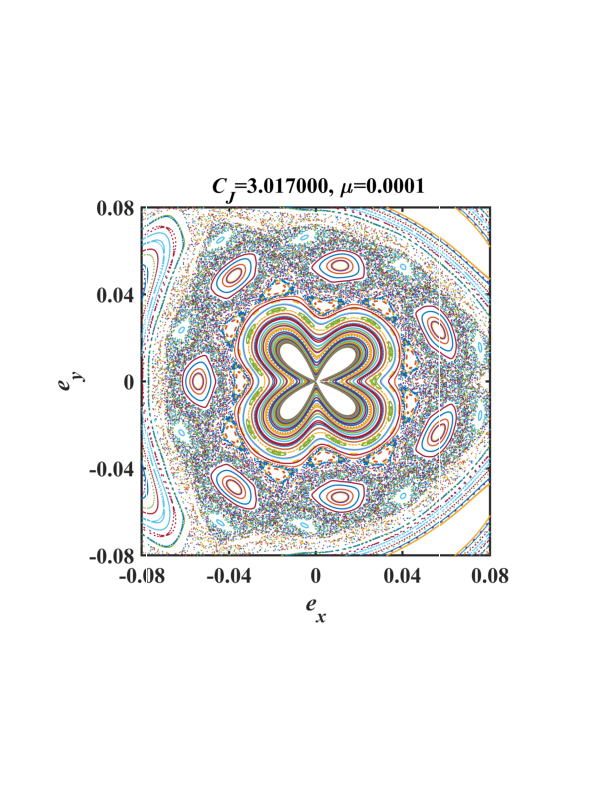}
 	\includegraphics[width=0.33\textwidth]{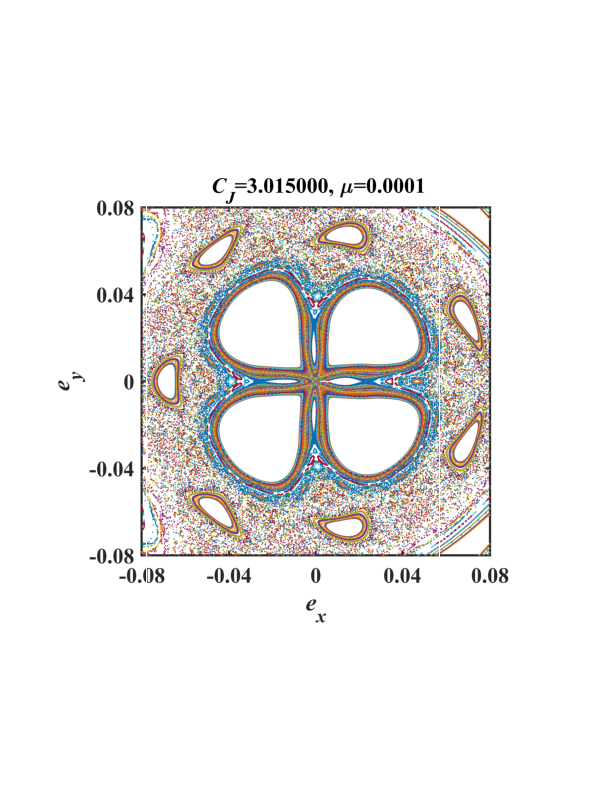}
  	\includegraphics[width=0.33\textwidth]{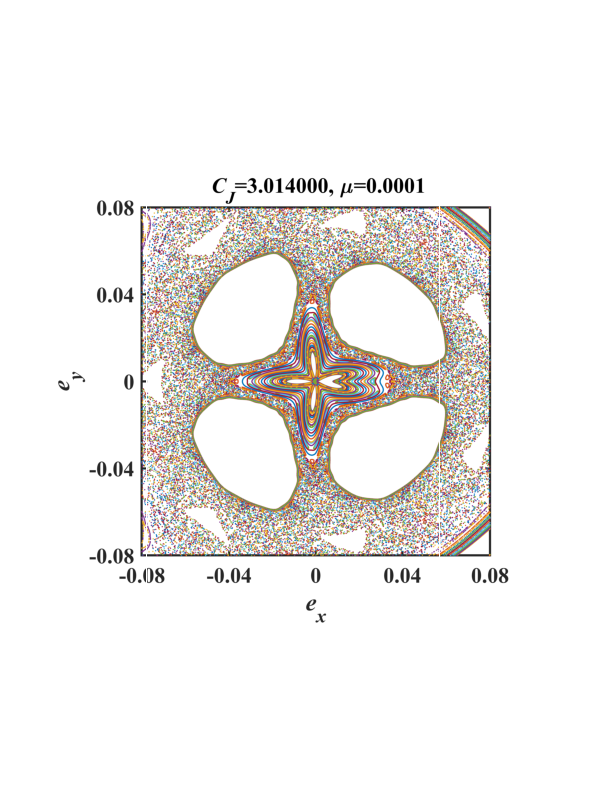}
   \vskip-0.5truein
    \caption{Poincar\'e surfaces-of-section for the 4:5 exterior MMR for $\mu=0.0001$.}
    \label{fig:fig1}
\end{figure*}

\begin{figure*}
\vskip-0.5truein
	\includegraphics[width=0.33\textwidth]{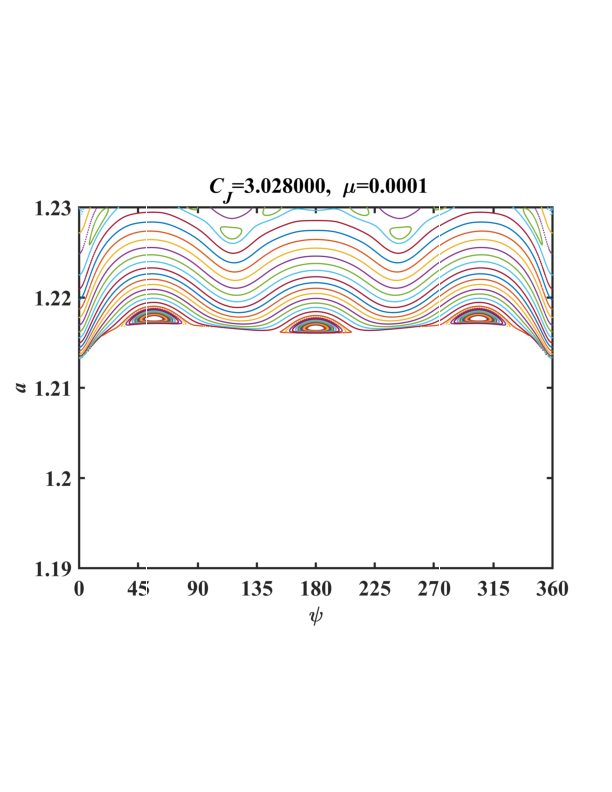}
 	\includegraphics[width=0.33\textwidth]{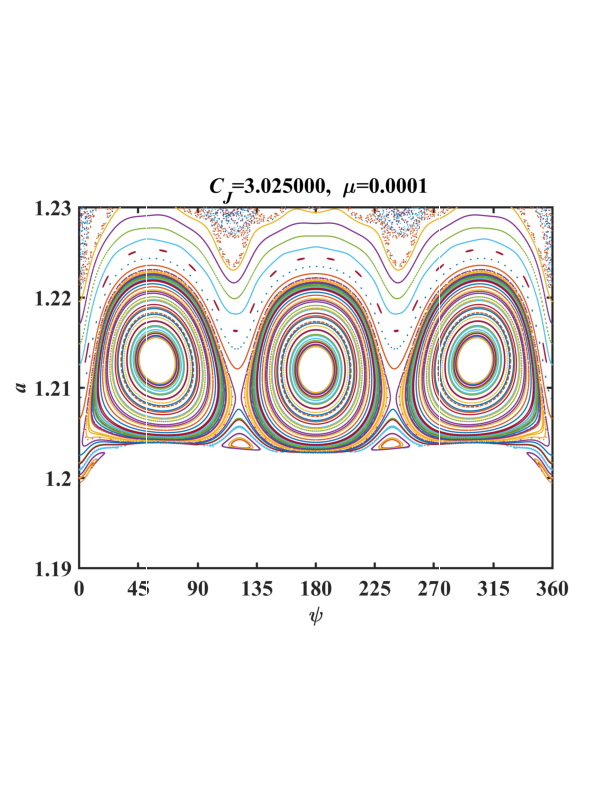}
  	\includegraphics[width=0.33\textwidth]{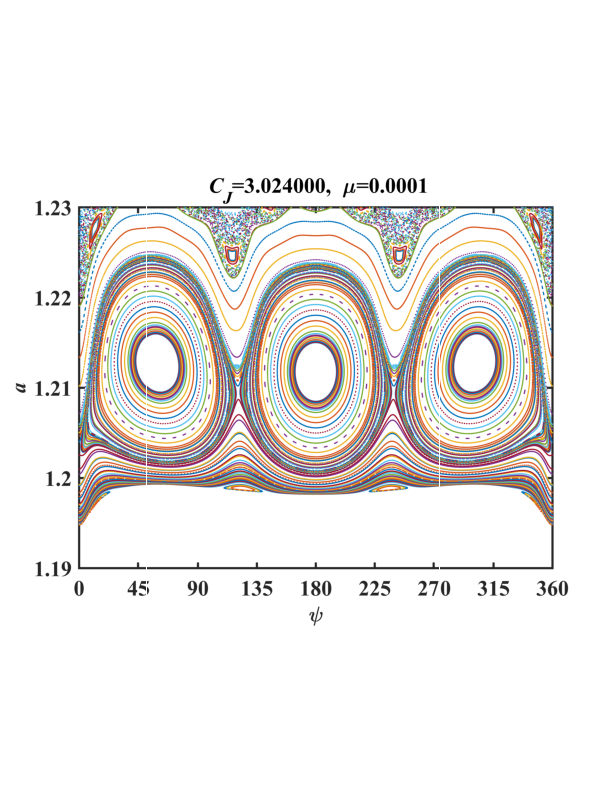}\\
   \vskip-1.1truein
    \includegraphics[width=0.33\textwidth]{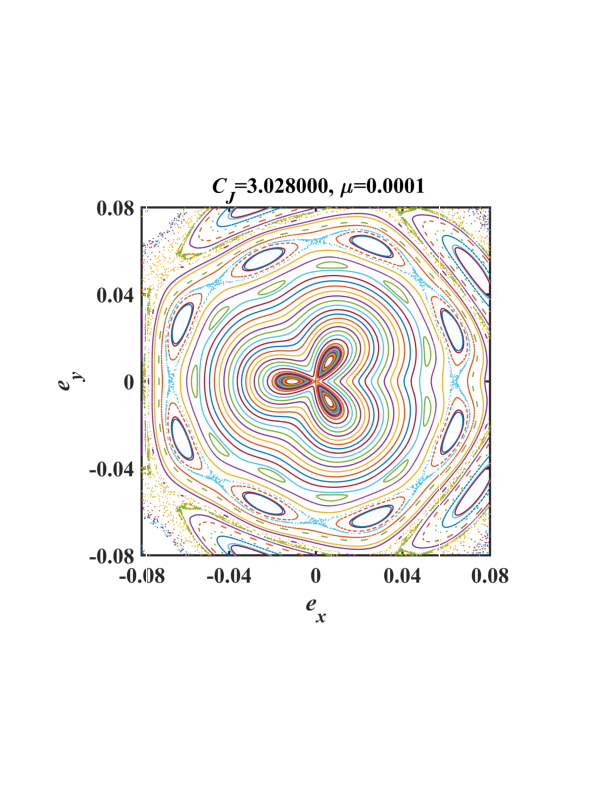}
 	\includegraphics[width=0.33\textwidth]{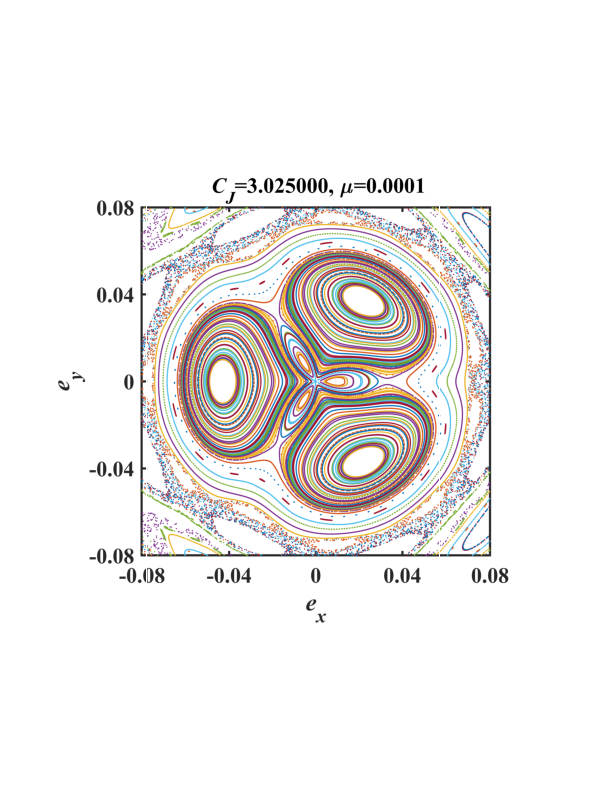}
  	\includegraphics[width=0.33\textwidth]{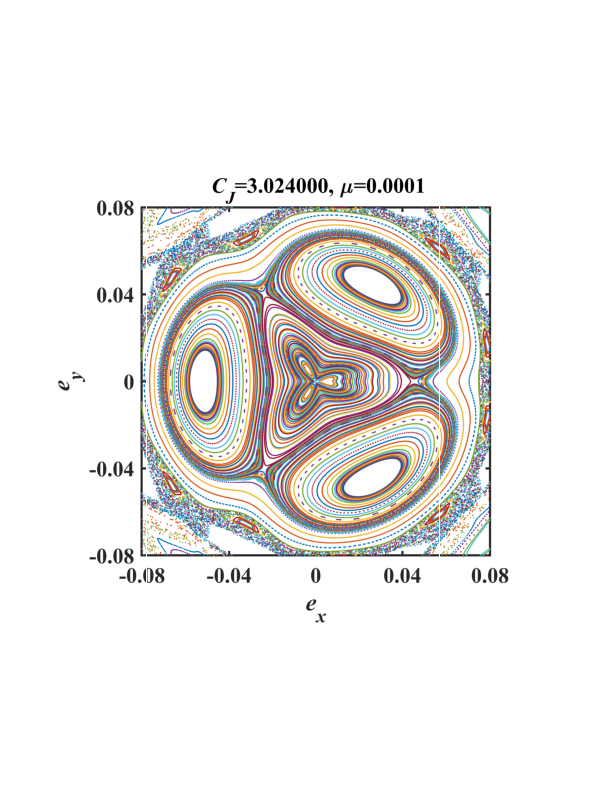}  
    \vskip-0.5truein
    \caption{Poincar\'e surfaces-of-section for the 3:4 exterior MMR for $\mu=0.0001$.}
    \label{fig:fig2}
\end{figure*}

\begin{figure*}
    \vskip-0.5truein
	\includegraphics[width=0.33\textwidth]{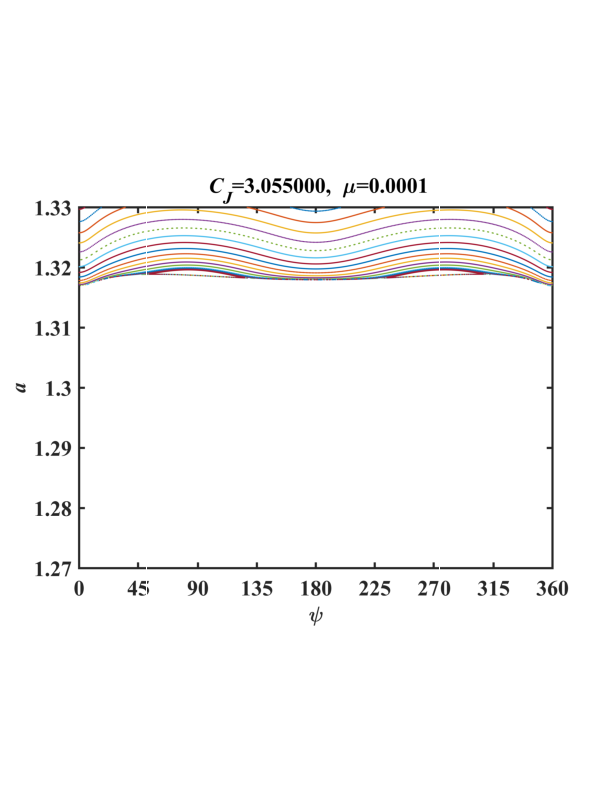}
 	\includegraphics[width=0.33\textwidth]{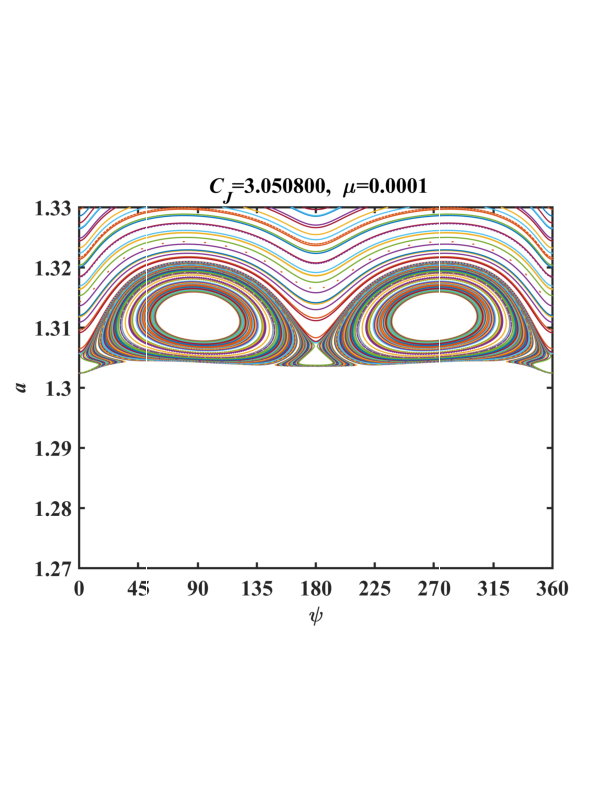}
  	\includegraphics[width=0.33\textwidth]{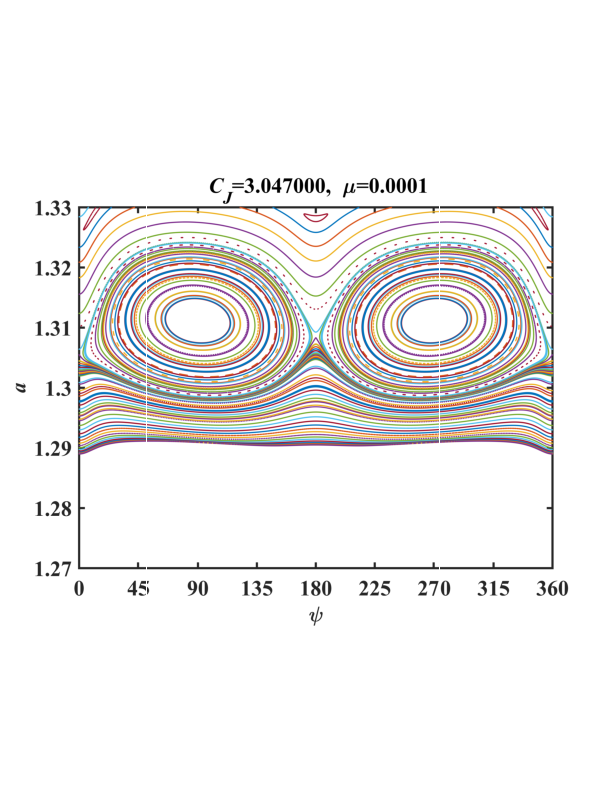}\\
   \vskip-1.1truein
    \includegraphics[width=0.33\textwidth]{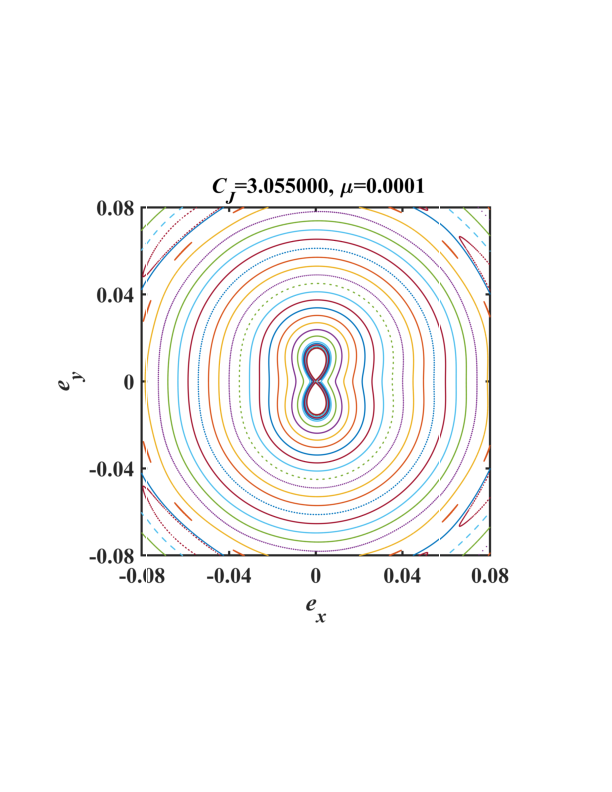}
    \includegraphics[width=0.33\textwidth]{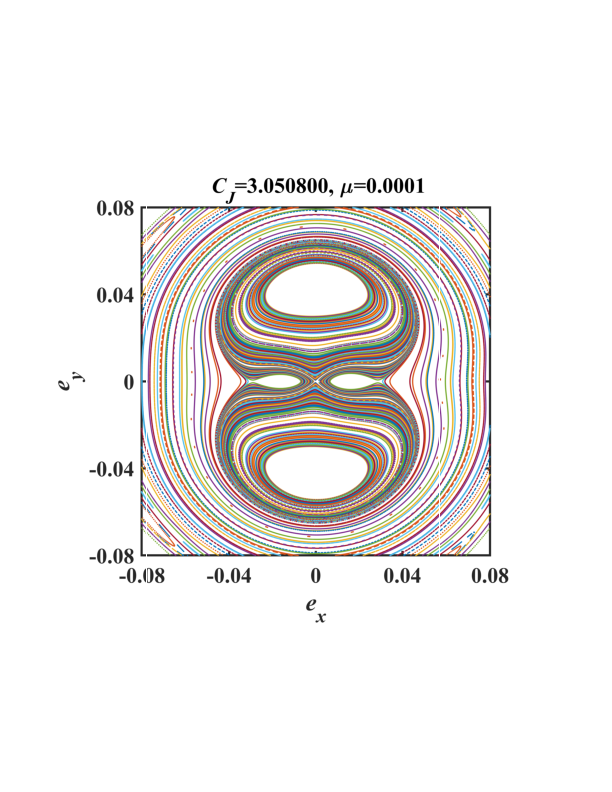}
    \includegraphics[width=0.33\textwidth]{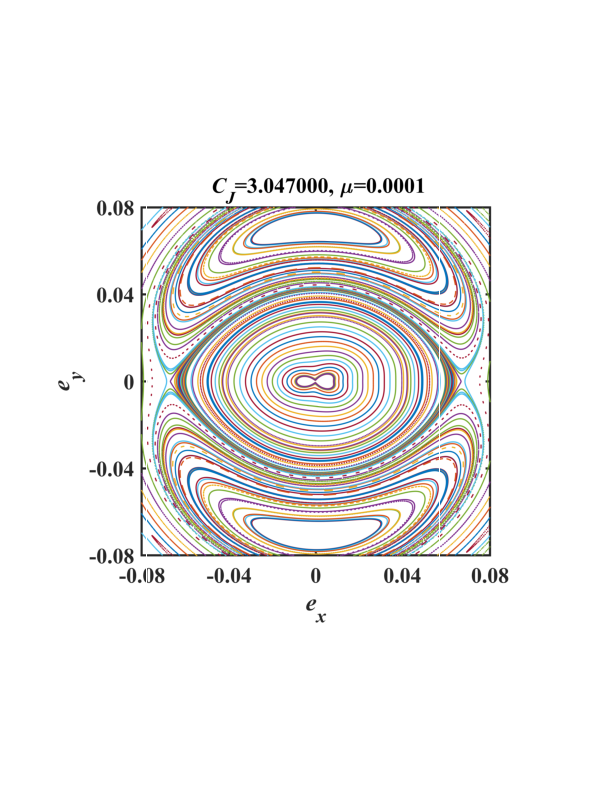}
    \vskip-0.5truein
    \caption{Poincar\'e surfaces-of-section for the 2:3 exterior MMR for $\mu=0.0001$.}
    \label{fig:fig3}
\end{figure*}

\begin{figure*}
    \vskip-0.5truein
	\includegraphics[width=0.33\textwidth]{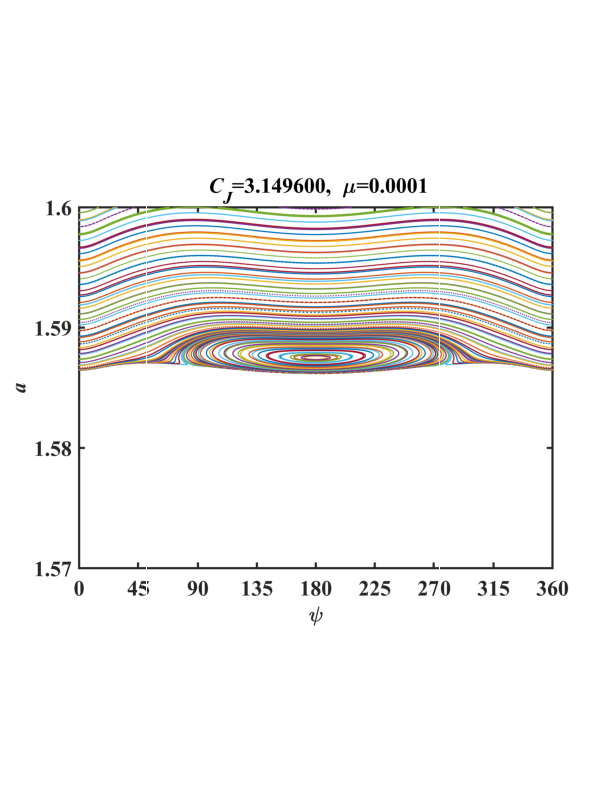}
 	\includegraphics[width=0.33\textwidth]{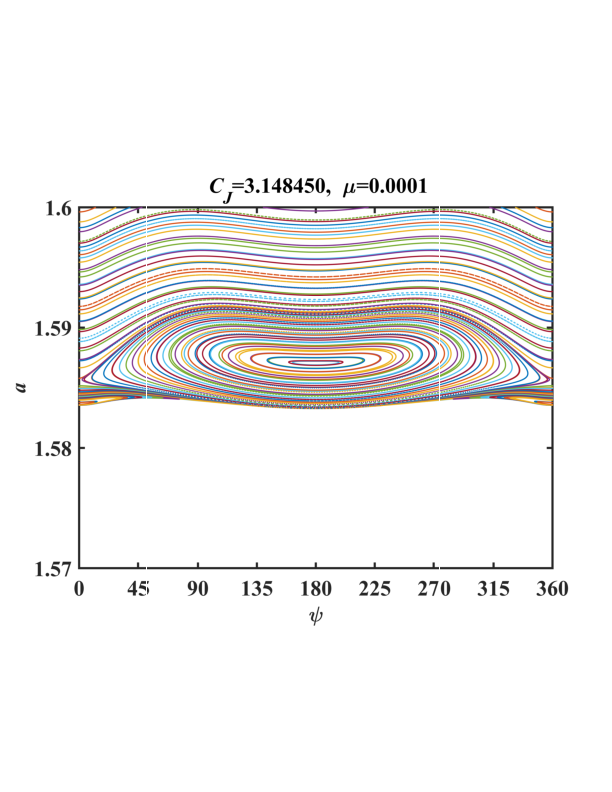}
  	\includegraphics[width=0.33\textwidth]{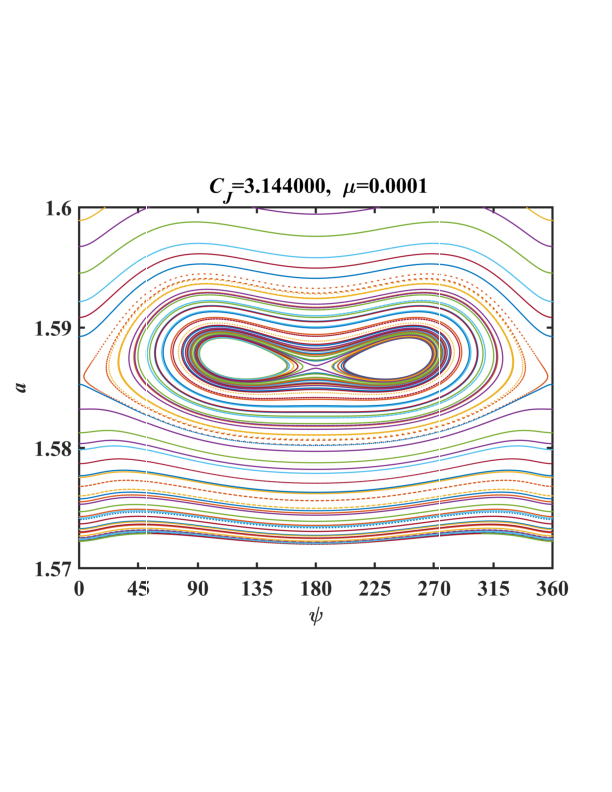}\\
   \vskip-1.1truein
    \includegraphics[width=0.33\textwidth]{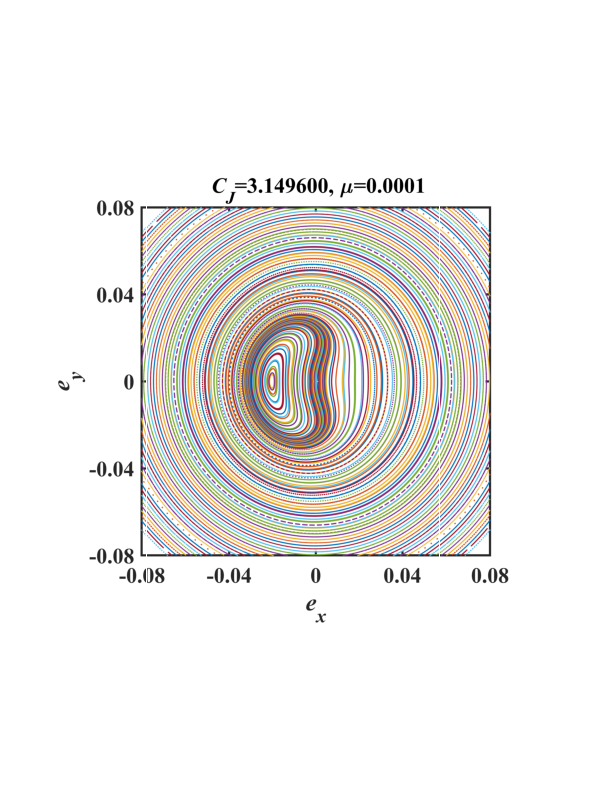}
    \includegraphics[width=0.33\textwidth]{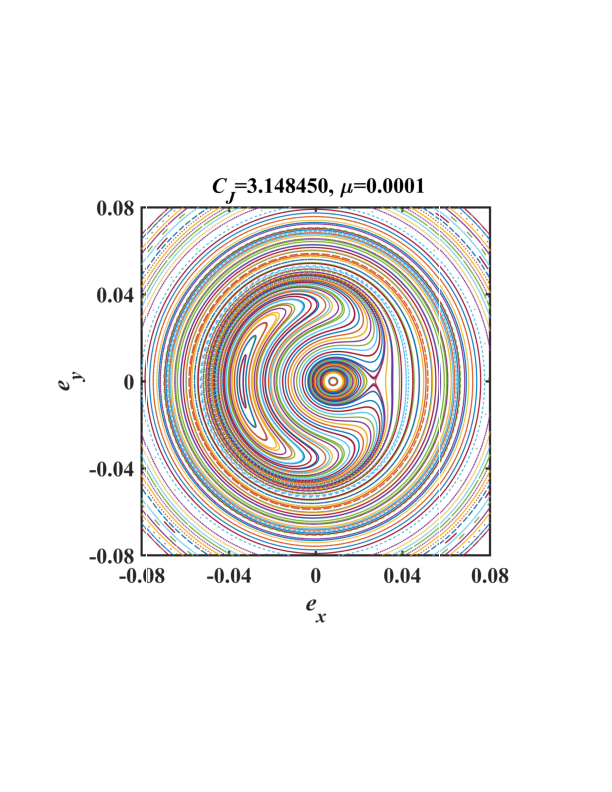}
    \includegraphics[width=0.33\textwidth]{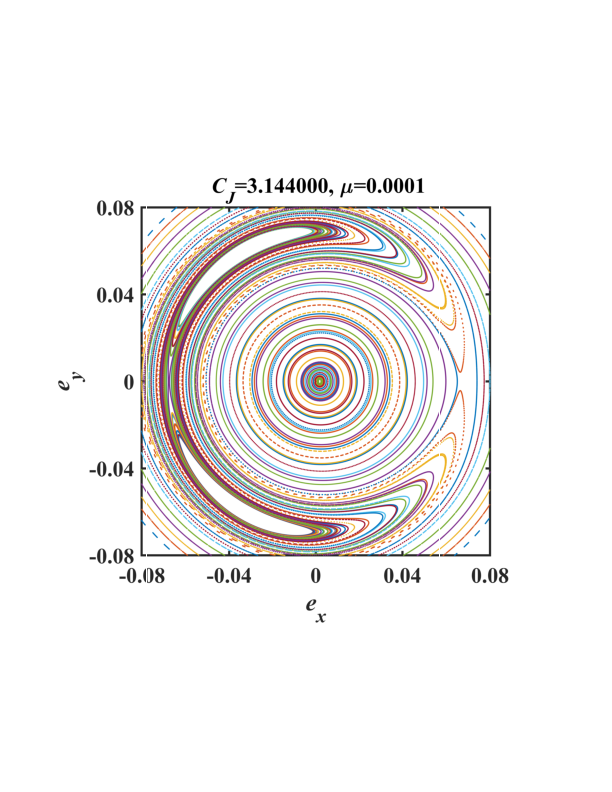}
    \vskip-0.5truein
    \caption{Poincar\'e surfaces-of-section for the 1:2 exterior MMR for $\mu=0.0001$.}
    \label{fig:fig4}
\end{figure*}

\begin{figure*}
	\hglue-0.3truein
    \includegraphics[width=1.1\textwidth]{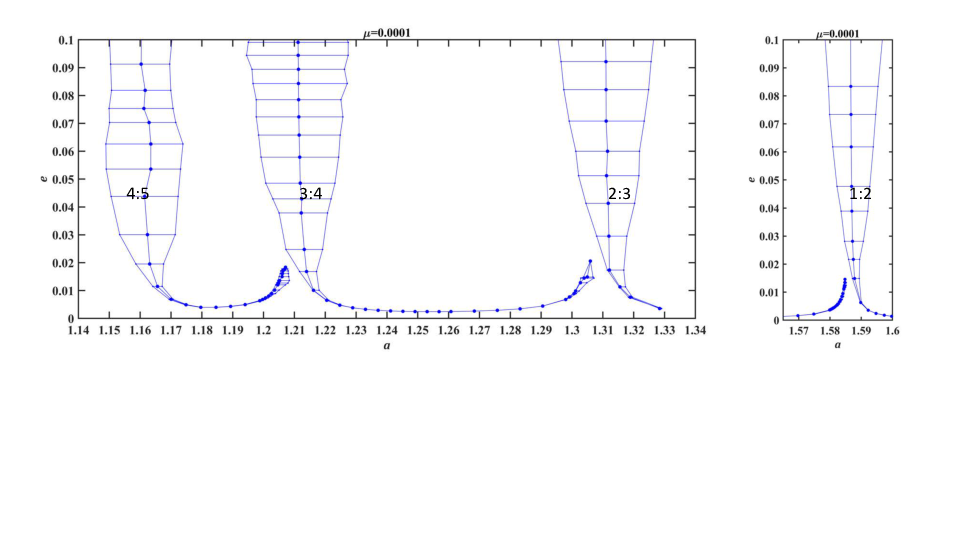}
	\vglue-1.4truein
	\caption{First order resonance widths in the $(a,e)$ plane, for $\mu=0.0001$. }
    \label{fig:fig5}
\end{figure*}

In each of Figures \ref{fig:fig1}--\ref{fig:fig4}, we plot Poincar\'e surfaces-of-section for the first order exterior mean motion resonances, 4:5, 3:4, 2:3 and 1:2, respectively, for the case of $\mu=1\times10^{-4}$. Each figure shows surfaces-of-section for three different values of the Jacobi constant, $C_J$, to illustrate how the phase space changes with $C_J$. The three values of $C_J$ are in descending order, and generally correspond to increasing values of $e_\mathrm{res}$ within each resonance. We observe that, in these Poincar\'e sections, a $j:(j+k)$ exterior MMR presents as a chain of $j$ resonant islands with approximate $j$--fold symmetry. In the $(\psi,a)$ plots, the visible lower boundary of the trajectories is enforced by the value of the Jacobi constant (noted in the top legend of each plot); below that boundary, there are no solutions for that specific value of the Jacobi constant. The most prominent chain of $j$ islands visible in these plots is that of the apocentric librations in which the particle's conjunctions with the planet librate around its apocenter. At the lower values of $C_J$, we observe an additional chain of $j$ resonant islands nested around the apocentric islands; these are most readily visible very close to the origin in the $(e_x,e_y)$ plots. This is the chain of islands of the pericentric librations, and they exist only for very small eccentricities, below a maximum value of a few percent. 

In the case of the 1:2 exterior MMR, at very low eccentricity, there is only one resonant island for the apocentric zone and one for the pericentric zone. However, at slightly higher eccentricities, the apocentric zone supports two asymmetric libration zones embedded within a symmetric libration zone (see the third pair of panels Figure~\ref{fig:fig4}). The asymmetric libration zones arise from a bifurcation of the apocentric resonance center. As mentioned in Section~\ref{s:intro}, this is a feature peculiar to exterior $1:j$ MMRs.

The $j$--fold symmetry of the chains of resonant islands is only approximate. This can be understood by considering the system in the rotating frame.  Recall that, in the rotating frame, the massive bodies are at fixed locations on the $x$--axis. Their fixed locations are not symmetric about the origin (except for the unique case of equal mass primaries). When we measure the state vector of the particle at pericenter, the angular phase $\psi$ is measuring the polar angle of the pericenter from the $x$-axis in the rotating frame. We can expect that the phase space portraits  will have symmetry about the $x$-axis but not about the $y$-axis.  Consequently, only those resonant islands that are located symmetrically about the $x$--axis have symmetry under the transformation $y\longrightarrow -y$. This means that their centers and widths will be identical when measured in terms of their osculating semimajor axis and eccentricity. For example, in Fig.~\ref{fig:fig1}, the pair of islands centered at $\psi\approx45^\circ$ and $\psi\approx315^\circ$ have common values of $a_\mathrm{res}$ and $e_\mathrm{res}$ as well as the range $a_\mathrm{min},a_\mathrm{max}$ of their extent. Similarly the pair of islands centered at $\psi\approx135^\circ$ and $\psi\approx225^\circ$ have common values of $a_\mathrm{res}$ and $e_\mathrm{res}$ as well as the range $a_\mathrm{min},a_\mathrm{max}$ of their extent. However, the values for the latter pair differ slightly from that of the former pair, even though both pairs belong to the same resonance. These differences are visible by eye in the $(\psi,a)$ and $(e_x,e_y)$ plots. 

In Figure~\ref{fig:fig5}, we plot the resonance centers, $(a_\mathrm{res},e_\mathrm{res})$, and the resonance widths in the $(a,e)$ parameter plane for several first order MMRs, 4:5, 3:4, 2:3 and 1:2, for perturber mass fraction $\mu=0.0001$. (Other cases of $\mu$ are generally similar.) We observe two features at low eccentricities:  each MMR is split into two branches, and there exist bridges between neighboring first order MMRs. For each MMR, the dominant visible resonance branch in this figure is the apocentric resonance zone in which the particle's conjunctions with the planet librate about its apocenter. The second branch, visibly smaller, is the pericentric resonance zone in which the particle's conjunctions with the planet librate about its pericenter. As mentioned previously, this branch exists only for very small eccentricities, below a maximum value of a few percent. Indeed, the pericentric zone of the 4:5 MMR was too small to measure by eye, so we limited our measurements of the pericentric branches to the 3:4 and more distant first order MMRs.

At very small eccentricities the center of each branch moves away from the nominal location of the MMR [i.e., away from $a^*_\mathrm{res}=((j+1)/j)^\frac{2}{3}$]. For decreasing values of $e_\mathrm{res}$, the center of the apocentric zone moves to the right, that is, to higher values of the semimajor axis, whereas the center of the pericentric zone moves to the left, to smaller values of semimajor axis. However, far away from the nominal resonance location, the eccentricity does not asympotically approach zero, as is found in perturbative approaches \citep[e.g.,][]{Henrard:1983}. Instead, accompanying the migration of the resonant center, we observe that the pericentric zone of the $j:(j+1)$ MMR smoothly transforms into the apocentric zone of the $(j+1):(j+2)$ MMR, making the ``low eccentricity resonant bridge" between these neighbor first order MMRs. The width of the libration zone in this bridge is much smaller than the width of the apocentric resonance zone closer to the nominal semi-major axis location of the MMR. 

By examining the Poincar\'e sections across these resonant bridges, we observe two noteworthy features. The first is that the boundaries of the pericentric islands in the Poincar\'e sections in the eccentricity plane, $(e_x,e_y)$, pass through a common point at zero eccentricity. This indicates that the zero eccentricity orbit is a homoclinic orbit, and that the phase space in this region sports a separatrix that passes through zero eccentricity. The zero eccentricity orbit is an unstable periodic orbit in this range of semimajor axis that forms the resonant bridge. 

The second feature we observe is that the transformation from the pericentric zone of the $j:(j+1)$ MMR to the apocentric zone of the $(j+1):(j+2)$ MMR involves a bifurcation of one of the $j$ pericentric resonant islands, the one centered at $\psi=0$. An example of this incipient bifurcation is visible in the lower right panel in Fig.~\ref{fig:fig1} where we see that the pericentric island (of the 4:5 MMR) centered at $\psi=0$ has a nearly bifurcated shape, visibly different from the same island in the lower middle panel at a slightly higher value of $C_J$. (A similar feature is visible in the lower right panel of Fig.~\ref{fig:fig2} as well as of Fig.~\ref{fig:fig3}, for the 3:4 and 2:3 MMRs, respectively.) With this bifurcation, the $j$ pericentric islands transform into $(j+1)$ islands as $a_\mathrm{res}$ moves to lower values along the resonant bridge. Simultaneously, the other islands in the resonant chain\footnote{The terminology "resonant chain" we use here refers to the structures in the Poincare' sections; "resonant chain" is sometimes also used to refer to cases of "three-body resonances” in systems of three (or more) planets when more than one pair of planets are in or near resonance.} gradually move their centers to positions that have approximate $(j+1)$-fold symmetry.

The zero eccentricity separatrix and the ``low eccentricity resonant bridges" were previously reported for the first order interior MMRs of Jupiter \citep{Malhotra:2020}. Our results here demonstrate that they also exist for first order exterior MMRs.

\begin{figure*}
    \hglue-0.5truein
	\includegraphics[width=1.15\textwidth]{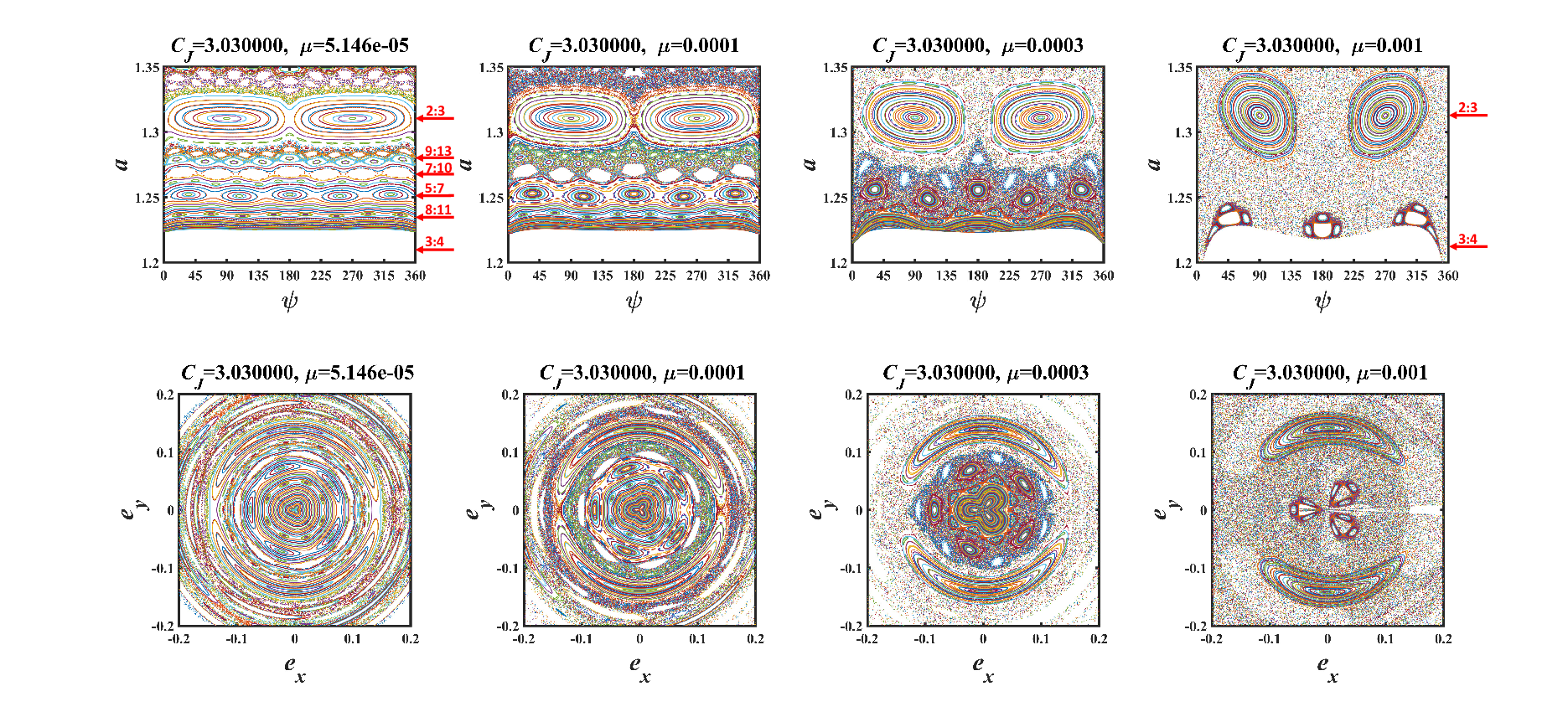}
	\vglue-0.0truein
	\caption{A sequence of Poincar\'e surfaces of section for increasing planet mass, $\mu$, for the same Jacobi constant, $C_J=3.03$.}
    \label{fig:fig6}
\end{figure*}

In Figure~\ref{fig:fig6}, we plot a sequence of Poincar\'e surfaces-of-section for different values of the planet mass, $\mu$, but at a fixed value of the Jacobi constant, $C_J=3.03$. This figure illustrates the $\mu$--dependence of the phase space structure around the 2:3 exterior resonance. For the lowest mass ratio, $\mu=5.146\times10^{-5}$, we observe that the 2:3 MMR has a well-resolved boundary with no visible chaotic sea. Moreover, many other chains of resonant islands are cleanly resolved; we can identify these structures corresponding to several higher order MMRs, such as 9:13, 7:10, 5:7, 8:11, 3:4, as indicated in the figure. For the larger values of $\mu$, we observe that more and more of these higher order resonant chains disappear into increasing amounts of area occupied by a chaotic sea. For the largest mass ratio that we investigated, $\mu=10^{-3}$, the higher order resonances are no longer discernible in Fig.~\ref{fig:fig6}: only the two islands of the 2:3 and the three islands of the 3:4 MMRs are visible, and these are surrounded by a vast chaotic sea. We also observe that, with increasing $\mu$, the stable domains of first order MMRs have larger widths in $a$ and $e$ but smaller widths in libration amplitude of $\psi$. Because the resonance width in $a$ and $e$ translates into the radial width of the resonance and the width in $\psi$ (equivalently, $\phi$) translates into the azimuthal width of the resonance, our results show that for larger $\mu$ first order MMRs have larger radial extent but smaller azimuthal extent of the stable resonant librations. 

\begin{figure*}
    \vglue-0.5truein
	\includegraphics[angle=90,width=\textwidth]{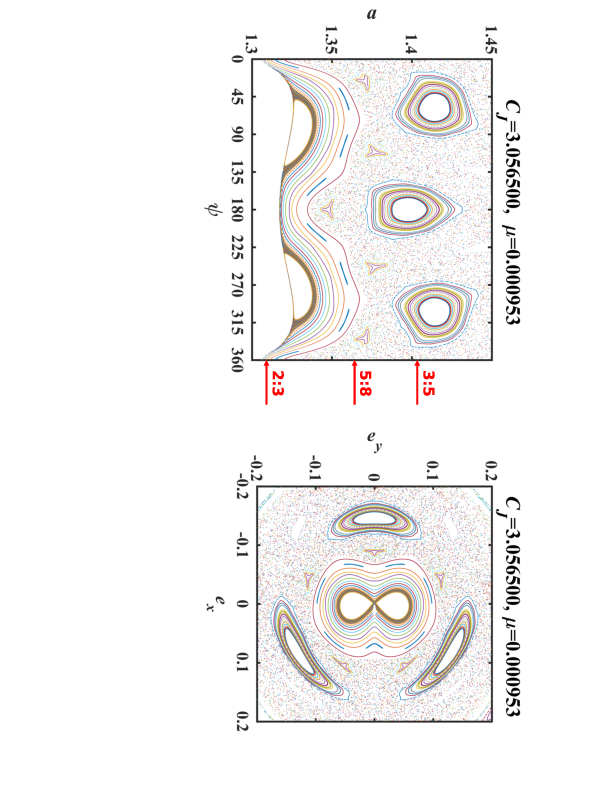}
	\vglue-1.7truein
    \caption{Poincar\'e surface-of-section to illustrate that neighboring mean motion resonances are nested in phase space but overlap in semimajor axis. The nominal locations, $a^*_\mathrm{res}$, of a few MMRs are marked.
    }
    \label{fig:fig7}
\end{figure*}

\begin{figure*}
	\includegraphics[width=\textwidth]{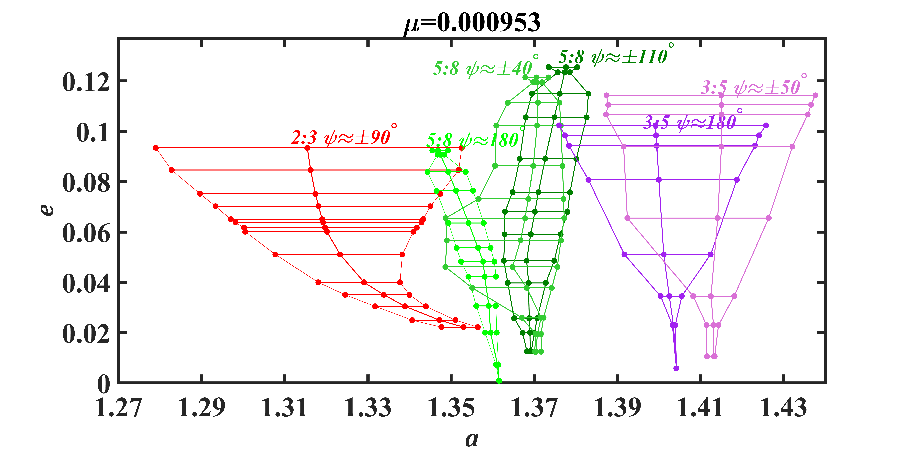}
    \caption{Neighboring mean motion resonances in the $(a,e)$ plane that overlap in semimajor axis but occupy distinctly separate regions in phase space (see Fig.~\ref{fig:fig7}). The location and extent of the 2:3 MMR is shown in red, the locations and extent of different islands of the 5:8 MMR are shown in different shades of green, and the locations and extent of the different islands of the 3:5 MMR are shown in shades of violet.}
    \label{fig:fig8}
\end{figure*}

In Figure~\ref{fig:fig7}, we illustrate the interesting fact that higher order resonances are intricately ``nested" in some parts of phase space. We observe that the chain of five islands of the 5:8 MMR is nested in-between the chain of three islands of the 3:5 MMR and the two-island chain of the 2:3 MMR. When we plot these resonance zones in the $(a,e)$ plane, as in Figure~\ref{fig:fig8}, they overlap in projection, even as they exist as distinct and non-overlapping libration islands in the higher dimensional phase space. This underscores an important point: the commonly used concept of ``resonance overlap" must be understood with this nuance that stable resonant librations can exist even when there is overlap of resonance widths in the semimajor axis range (or, equivalently, in the range of orbital period ratios).

\begin{figure*}
	\vglue-0.5truein
    \includegraphics[angle=90, width=1.15\textwidth]{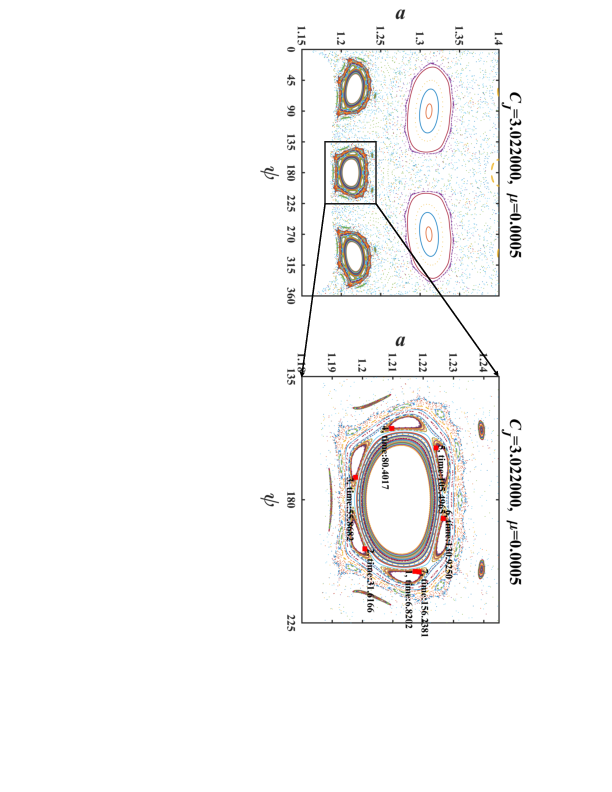}
    \vglue-2.5truein
    \caption{A Poincar\'e surface of section displaying a chain of secondary resonance islands in the 3:4 MMR. The panel on the right illustrates the sequence of visits of one trajectory librating in the 3:4 MMR close to this chain of six secondary resonance islands.}
    \label{fig:fig9}
\end{figure*}

In Figure~\ref{fig:fig9} we illustrate the phenomenon of secondary resonances. These are chains of islands that surround the main islands of the mean motion resonances. We detected secondary resonance chains around many low order MMRs, most commonly for the higher mass perturbers of $\mu\ge5\times10^{-4}$. Such secondary resonances are well known in nonlinear dynamical systems \citep[e.g.,][]{Chirikov:1979}. They arise from the interaction of the main resonance with other frequencies in the perturbing potential \citep[e.g.,][]{Tittemore:1989,Malhotra:1990c}. 
{Focusing on the 3:4 MMR resonant island centered at $\psi=180^\circ$ (Figure~\ref{fig:fig9}), we can identify a prominent chain of six islands (highlighted in the exploded panel on the right). Moving outward to larger libration amplitude, we can also readily identify a chain of 17 islands, a chain of 11 islands and a chain of 5 islands; the keen reader may notice additional chains. [Note that the secondary resonance islands occur within the libration zones of the main MMR islands and they encircle the center of each of the main MMR islands; they are distinctly different from the chains of islands of MMRs, including higher order MMRs, that appear as  nearly horizontal structures in the $(\psi,a)$ plots, such as those marked in Figure~\ref{fig:fig6}, top left panel.] By measuring the periods of the trajectories near these secondary resonance chains of islands, we found that these are owed to commensurability of the libration frequency of the main resonance with the synodic frequency, the latter being the difference of the orbital frequencies of the planet and the particle.  To illustrate, we observe in Figure~\ref{fig:fig9} (right panel) that the libration period of a trajectory belonging to the 6-island secondary resonance chain is $\Delta t \approx 156.2381-6.8202 = 149.4179$ (in natural units, where the orbital period of the primaries is $2\pi$); this is the time interval between point \#1 and point \#7 which (nearly) completes a full libration of $\psi$. Therefore, the libration frequency of trajectories near this secondary resonance is $\omega_\mathrm{lib}\approx 2\pi/149.4179 \approx0.04205$. The orbital frequency of the particle at the center of the main 3:4 resonance in this case is $n\approx1.214^{-\frac{3}{2}}\approx0.7476$, and the orbital frequency difference between the planet and the particle is $\Delta n \approx 0.2524$, so that $|6\times\omega_\mathrm{lib}-\Delta n|/\Delta n \approx 4\times10^{-4}$. These calculations show that the libration frequency near the chain of six secondary resonance islands has a 6/1 commensurability with the synodic frequency.  In the Appendix, we describe additional examples of secondary resonances. In general, a chain of $N$ secondary resonance islands is associated with an $N/k$ commensurability between the libration frequency and the synodic frequency, where $N$ and $k$ are mutually prime numbers and $k<N$.}

At the same value of $C_J$ as in Figure~\ref{fig:fig9}, but at a higher perturber mass, $\mu=9.53\times10^{-4}$, we detected only a chain of four secondary resonance islands, i.e., a 4/1 commensurability of the libration frequency with the synodic frequency. % slide #25 in Work Report.pptx
{No other secondary resonance chains appear in this case, and the range of $\psi$ of the main resonance zone is smaller. This indicates that for larger mass perturbers, the higher integer secondary resonances (of the type $N/k$, with larger $N$) are increasingly submerged in the chaotic sea. The shrinking of the libration amplitude and the increase of the chaotic sea surrounding the main resonance zones can be attributed in part to the merging of secondary resonances at larger libration amplitudes on the shores of the main resonance islands. }

{The phenomenon of secondary resonances deserves some elaboration. Because the libration frequency depends on libration amplitude, as well as on $\mu$ and on $e$, in a complex way, the appearance and locations of secondary resonances is quite complex. For example, \cite{Malhotra:1996} and \cite{Lan:2019} computed the libration period as a function of libration amplitude for a few exterior MMRs of Neptune, and for a few different values of the eccentricity, using non-perturbative approaches. For low eccentricities, their results showed that the libration period in the 1:2 and 2:3 MMRs changes very slowly at first with increasing libration amplitude, then changes rapidly and reaches a maximal (very large) value over a small range of high libration amplitude (e.g., Fig.~3 and Fig.~6 of \cite{Malhotra:1996}), then decreases sharply again (e.g., Fig.~4 and Fig.~11 of \cite{Lan:2019}). The rapid variation of the libration period at large libration amplitudes implies that many secondary resonances, with higher values of $N$, occur in this region of the main resonance zone. And these occur in close proximity to each other, as seen in Figure~\ref{fig:fig9}. Consequently, at large libration amplitudes, the density of secondary resonances will become high enough that their overlap will lead to chaos, shrinking the stable domain of the main resonant islands from the inside-out. Only the strongest of the secondary resonances are visible as chains of islands, the others are submerged in the chaotic sea.}

\subsection{Comparison with previous results}

{The literature cited in Section 1 describes many of the results from the long history of previous theoretical investigations of mean motion resonances. Here we limit to a comparison of our results with those from \cite{Lei:2020}'s perturbative analysis of several of the same MMRs we investigated here. These authors adopted the single resonance approximation and a disturbing function (for the circular planar restricted three body model) truncated up to order 10 in powers of eccentricity to investigate several first order inner and outer mean motion resonances. They reported results for a single value of the mass fraction, $\mu$ corresponding to the Jupiter-Sun system, and compared their results to the non-perturbative results of \cite{Malhotra:2020}. They confirmed that the new critical resonant angle $\phi/j$ (see Eq.~\ref{e:phi}), introduced in \cite{Wang:2017}, has the advantage of making visible important phase space structures that are suppressed in the analyses with the traditional critical resonant angle, $\phi$. Notably, previous work has nearly exclusively made use of only $\phi$; this is probably because this angle arises in series expansions of the disturbing function, whereas its sub-multiple, $\psi$, does not. 

\cite{Lei:2020} found that the topology of the phase space can be erroneous when computed with only one or two lowest order harmonics in the disturbing function, but can be more accurately recovered by including higher order harmonics. One significant feature of the phase space topology that is not recovered with only one or two lowest order harmonics is the existence of the separatrix passing through zero eccentricity. We can observe that the topology of the phase space of the 3:4 and the 2:3 exterior MMRs (Poincar\'e sections in the $(e_x,e_y)$ plane in our Fig.~2 and Fig.~3) is quite similar to \cite{Lei:2020}'s Fig.~C3 and Fig.~8; those authors did not report phase space portraits for the exterior 1:2 MMR, so we cannot compare with our Fig.~4 for this MMR. 

We also observe some differences between \cite{Lei:2020}'s perturbative analyses and our non-perturbative analyses: (a) The former does not make visible the secondary resonances and the chaotic seas that exist near the boundaries of the stable libration regions. (b) The resonant bridges between neighboring first order MMRs are also not visible with the perturbative approach. (c) The gradual transformation of the phase space topology from approximate $j$-fold symmetry to approximate $(j+1)$-fold symmetry (and vice versa) at low eccentricities away from the nominal resonance locations is also not visible in \cite{Lei:2020}'s results. These differences are due to the single resonance assumption in the perturbative analysis, as this assumption neglects the effects of neighboring MMRs as well as the synodic perturbations from the planet; the non-perturbative analysis includes these effects fully without approximations. 
}

\section{Summary and Discussion}\label{s:summary}

We investigated the phase space structure of exterior mean motion resonances in the regime of low eccentricities (up to about 0.1) with a non-perturbative, fully numerical approach using visualizations with Poincar\'e surfaces of section of the circular planar restricted three body model. We obtained results for a range of the perturber's mass fraction, $\mu$,  $\sim5\times10^{-5}$ to $1\times10^{-3}$ (corresponding to the mass ratio of Neptune-to-Sun and Jupiter-to-Sun). A summary of our findings as well as discussion of some of their phenomenological implications is enumerated below.
\begin{enumerate}
    \item We find that first order exterior MMRs have two branches at low eccentricities, the apocentric branch and a smaller pericentric branch, similar to the previous result for interior MMRs \citep[as in][]{Malhotra:2020}. It must be emphasized that this fine-scale structure is distinct from the well-known bifurcation of the apocentric resonance zone of $1:j$ exterior MMRs that occurs at higher eccentricities \citep{Beauge:1994,Malhotra:1996,Murray-Clay:2005,Lan:2019}.
    
    \item For decreasing eccentricity of the test particle, the center of the pericentric branch of a $j:(j+1)$ exterior MMR migrates away from the nominal resonance location, towards the $(j+1):(j+2)$ MMR location (closer to the perturber's orbit). The apocentric branch migrates in the opposite direction. However, in contrast with results from perturbative approaches \citep[e.g.,][]{Henrard:1983}, we find that the resonance center does not asympotically approach zero eccentricity at large distances from the nominal resonance location.
    
    \item Accompanying the migration of the resonant centers at very low eccentricities, the pericentric zone of the $j:(j+1)$ MMR smoothly transforms into the apocentric zone of the $(j+1):(j+2)$ MMR, generating a low-eccentricity bridge between neighboring first order MMRs.

    The low eccentricity resonant bridges are of great theoretical interest, but due to their narrow minimum widths, their role in nature remains to be investigated. \cite{Malhotra:2020} conjectured that these bridges may serve as effective transport conduits for radial migration under small non-gravitational forces (such as the Yarkovsky force due to solar radiation on small asteroids). \cite{Antoniadou:2021} confirmed this conjecture with numerical simulations, also noting that too slow radial migration could be interrupted by encounters with higher order MMRs. In a future investigation, this conjecture can be tested with observational data of small minor planets.
       
    \item With increasing mass fraction of the perturber, $\mu$, first order MMRs have larger widths in their radial extent (i.e., in $a$ and $e$) but smaller azimuthal widths (i.e., in the libration amplitude of their critical resonant angle, $\phi$). Moreover, with increasing $\mu$, more and more of the higher order resonances dissolve into the chaotic seas in the vicinity of lower order resonances. For Neptune-to-Sun value of $\mu$, we find that many MMRs are well resolved and have very little chaotic boundaries. However, for the Jupiter-to-Sun value of $\mu$ many high order MMRs dissolve into chaotic zones. This difference between Neptune's and Jupiter's resonances partially accounts for the observation that many of Neptune's exterior MMRs, including high order ones, are occupied by minor planets whereas Jupiter's higher order MMRs are mostly devoid of minor planets.

    \item Higher order resonances are often intricately ``nested" in some parts of phase space. This means that the commonly used concept of ``resonance overlap" causing chaos must be understood with some nuance, namely, that stable resonant librations can persist even when there is overlap of resonance widths measured in semimajor axis (or, equivalently, in terms of orbital period ratios).
    \item We detected many secondary resonances near the large libration amplitude boundaries of low order MMRs. We found that these arise from commensurabilities between the libration frequency of a main resonance and the synodic frequency with the planet. The implication is that secondary resonances are significantly responsible for shrinking the stable resonance widths and generating the chaotic sea surrounding MMRs of higher-mass perturbers. This result identifies a mechanism which is distinct from (and in addition to) the ``resonance overlap" mechanism for the origin of chaos, in which the interaction of neighboring MMRs is understood to be responsible for the dynamical chaos in the vicinity of MMRs. {In other words, the overlap of mean motion resonances can be understood to cause chaos ``from the outside-in" as neighboring MMRs press on each others' domain, while the secondary resonances cause chaos ``from the inside-out" as secondary resonances occur in the interior of the resonant libration zone.}
\end{enumerate}

\section*{Acknowledgements}

We thank the anonymous reviewer for comments that helped to improve this paper. RM thanks the Canadian Institute for Theoretical Astrophysics for hosting a sabbatical visit during the late stages of this work. ZC acknowledges research funding from Tsinghua University Initiative Scientific Research Program.

\section*{Data Availability}
{Table 1, Table A-1 and the figures contain the data generated in this work, and are available with the online edition of this article.}

%\bibliographystyle{mnras}
%\bibliography{allrefs} % if your bibtex file is called example.bib

%\clearpage
%\appendix

\section*{Appendix:
Additional examples of secondary resonances near the 3:4 exterior MMR}

\setcounter{figure}{0}                       
\renewcommand\thefigure{A-\arabic{figure}}
\setcounter{table}{0}                       
\renewcommand\thetable{A-\arabic{table}}

As noted in the main text, a chain of $N$ secondary resonance islands is associated with an $N/k$ commensurability between the libration frequency and the synodic frequency, where $N$ and $k$ are mutually prime numbers and $k<N$. An example is the 6/1 secondary resonance illustrated in Figure~\ref{fig:fig6}. Here we describe additional examples of secondary resonances in the same Poincar\'e section.

\begin{figure}
    \includegraphics[angle=90,width=0.95\columnwidth]{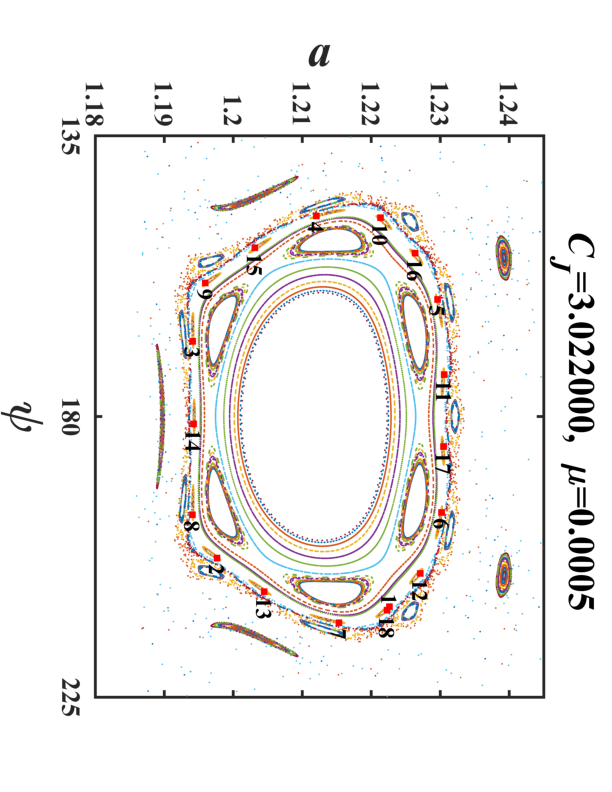}
    \caption{Similar to the right panel of Figure~\ref{fig:fig9}. This plot highlights the sequence of visits of one trajectory associated with the 17/3 secondary resonance within the exterior 3:4 MMR's resonant island centered at $\psi=180^\circ$.}
    \label{fig:figA1}
\end{figure}

In Figure~\ref{fig:figA1}, we highlight the 17-island chain. The sequence of visits in the Poincar\'e section of one quasi-periodic trajectory that belongs to this chain is marked by the red points, enumerated from 1 to 18. [These are the visits to the middle island (centered at $\psi=180^\circ$) of the three islands of the 3:4 MMR seen in the left panel of Figure~\ref{fig:fig9}. The time between successive visits to the Poincar\'e section is of course one orbital period of the particle, but the time between visits to any one of the three islands of the 3:4 MMR is three orbital periods of the particle, equivalently four orbital periods of the primaries, or approximately $8\pi$ in natural units.] The times of visit of each point in the sequence (taken from the numerical orbit integration) are tabulated in Table~\ref{t:tableA1}. Note that the successive points do not belong to the secondary resonance islands in sequence, rather in-between successive visits the trajectory skips over two of the islands in the secondary resonance chain. The 18th point returns close to the first one. During the interval between the first and 18th point the trajectory has wound three times around the center of the main island. This indicates that the libration period around the center of the 3:4 MMR is approximately one-third of the time interval between the 18th point and the first point. The numerical value of the 3:4 MMR libration frequency for this trajectory is therefore given by $\omega_\mathrm{lib}\approx (3\times2\pi)/(430.5514-6.9056) \approx0.04449$. The synodic frequency was noted previously in the main text, $\Delta n \approx 0.2524$. Therefore, $(17/3)\times\omega_\mathrm{lib} -\Delta n \approx 0.0011\Delta n $, showing that the libration frequency near the 17-chain secondary resonance has a 17/3 commensurability with the synodic frequency.

\begin{figure}
    \includegraphics[angle=90,width=0.95\columnwidth]{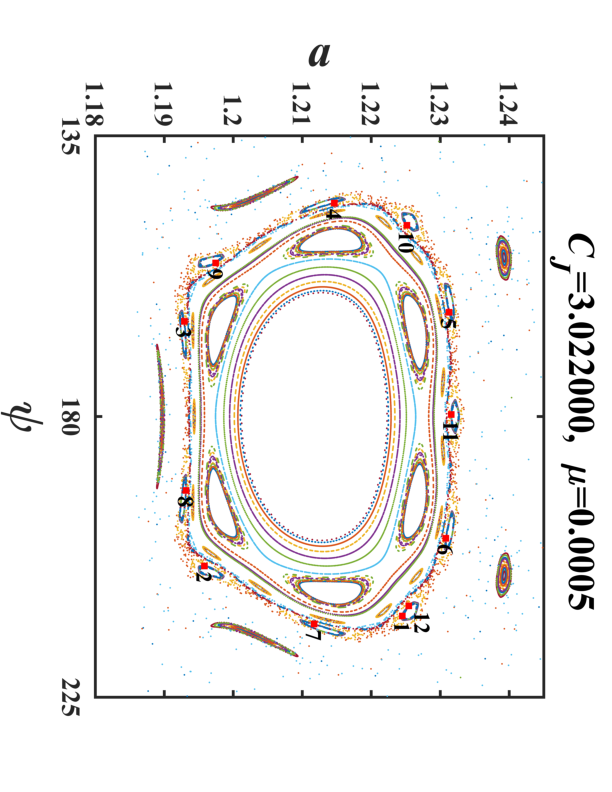}
    \caption{Similar to  Figure~\ref{fig:figA1}. This plot highlights the 11/2 secondary resonance.}
    \label{fig:figA2}
\end{figure}

\begin{figure}
    \includegraphics[angle=90,width=0.95\columnwidth]{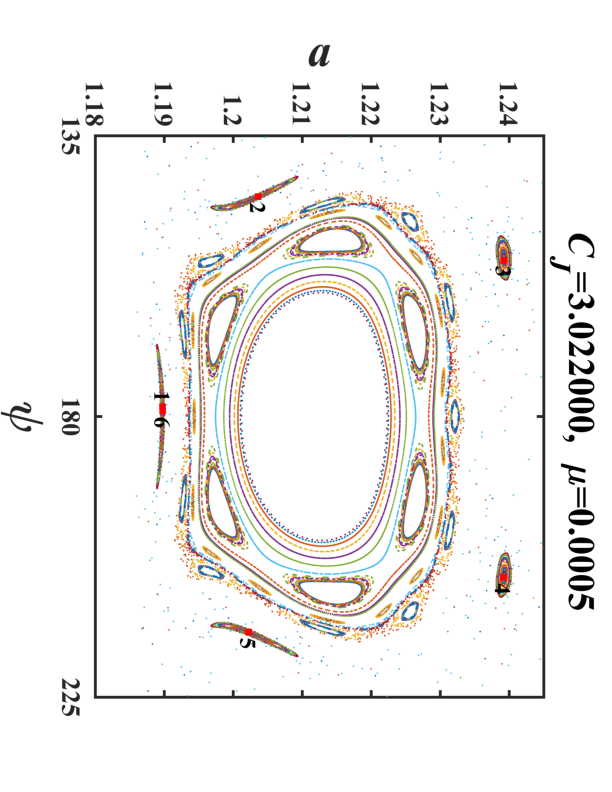}
    \caption{Similar to  Figure~\ref{fig:figA1}. This plot highlights the 5/1 secondary resonance.}
    \label{fig:figA3}
\end{figure}

Similarly, in Figure~\ref{fig:figA2} and Figure~\ref{fig:figA3}, we highlight the 11-island and the 5-island secondary resonance chains. The times of visit of each point in these chains (taken from the numerical orbit integration) are also tabulated in Table~\ref{t:tableA1}. In the case of the 11-island chain, in-between successive visits the trajectory skips over one of the islands in the secondary resonance chain. The 12th point returns close to the first one. During the interval between the first and 12th point the trajectory has wound two times around the center of the main island. This indicates that the libration period around the center of the 3:4 MMR is approximately one-half of the time interval between the 12th point and the first point. The numerical value of the 3:4 MMR libration frequency for this trajectory is therefore given by $\omega_\mathrm{lib}\approx (2\times2\pi)/(281.1520-6.9347) \approx0.04583$. Then using the synodic frequency, $\Delta n$, we find $(11/2)\times\omega_\mathrm{lib} -\Delta n \approx 0.0014\Delta n $, showing that the libration frequency near the 11-chain secondary resonance has a 11/2 commensurability with the synodic frequency.
In the case of the 5-island chain (visible within the large chaotic sea at the outskirts of the 3:4 MMR's main resonant islands), the 6th point returns close to the first one, and each of the 5 islands is visited in sequence, without any skips. The numerical value of the 3:4 MMR libration frequency for this trajectory is therefore given by $\omega_\mathrm{lib}\approx 2\pi/(131.0826-6.2306) \approx0.05033$. Then using the synodic frequency, $\Delta n$, we find $5\times\omega_\mathrm{lib} -\Delta n \approx 0.0030\Delta n $; we conclude that the libration frequency near the 5-chain secondary resonance has a 5/1 commensurability with the synodic frequency.

\begin{table}
\caption{
Data for successive points of three example trajectories forming chains of secondary resonance islands within the middle island of the 3:4 mean motion resonance (for $\mu=0.0005$). Taken from the same Poincar\'e section plotted in Figure~\ref{fig:fig9}. 
The time is given in natural units (see main text for explanation); an electronic version of this table, with double precision values, is available online.
}
\begin{center}
\begin{tabular}{clll}
\hline
        & 17-island chain & 11-island chain & 5-island chain\\
Point\# & Time  & Time & Time \\
\hline
1	&	6.9056	 &	6.9347	&	6.2306	\\
2	&	31.8313	    &	31.9473	&	30.3801	\\
3	&	55.8041	    &	55.7503	&	55.7469	\\
4	&	80.3727	    &	80.4354	&	81.6500	\\
5	&	105.6073	&	105.7439	&	107.0231	\\
6	&	131.1969	&	131.3724	&	131.0826	\\
7	&	156.5001	&	156.6239	&		\\
8	&	181.1950	&	181.1796	&		\\
9	&	205.0740	&	205.0948	&		\\
10	&	229.9702	&	230.1031	&		\\
11	&	255.3979	&	255.6247	&		\\
12	&	280.9451	&	281.1520	&		\\
13	&	306.0223	&		&		\\
14	&	330.3021	&		&		\\
15	&	354.5278	&		&		\\
16	&	379.5741	&		&		\\
17	&	405.1051	&		&		\\
18	&	430.5514	&		&		\\
\hline
\end{tabular}
\end{center}
\label{t:tableA1}
\end{table}

%\clearpage

% Don't change these lines
\bsp	% typesetting comment
\label{lastpage}
\end{document}